\documentclass[a4paper]{article}
\usepackage{Odyssey2020}
\usepackage{epsfig,amssymb,amsmath}
\usepackage[ruled,vlined]{algorithm2e}
\usepackage{setspace}
\usepackage{color}
\usepackage{cite}
\usepackage{soul}
\usepackage{comment}
\usepackage{multirow}
\ninept
\title{WaveTTS: Tacotron-based TTS with Joint Time-Frequency Domain Loss}

\name{Rui Liu$^{ 1,2}$ \thanks{\textbf{Speech Samples:} https://ttslr.github.io/WaveTTS/.}, Berrak Sisman$^{2,3}$, Feilong Bao$^{* 1}$, Guanglai Gao$^{ 1}$, Haizhou Li$^{ 2}$}

\address{$^1$ Inner Mongolia University, China \\
$^2$ Dept. of Electrical and Computer Engineering, National University of Singapore, Singapore \\ $^{3}$ Information Systems Technology and Design, Singapore University of Technology and Design \\
{\small \tt \{r.liu, berraksisman\}@u.nus.edu, \{csfeilong, csggl\}@imu.edu.cn,  haizhou.li@nus.edu.sg} }
\begin{document}
\maketitle
\begin{abstract}
Tacotron-based text-to-speech (TTS) systems directly synthesize speech from text input. Such frameworks typically consist of a feature prediction network that maps character sequences to frequency-domain acoustic features, followed by a waveform reconstruction algorithm or a neural vocoder that generates the time-domain waveform from acoustic features. As the loss function is usually calculated only for frequency-domain acoustic features, that doesn't directly control the quality of the generated time-domain waveform. To address this problem, we propose a new training scheme for Tacotron-based TTS, referred to as WaveTTS, that has 2 loss functions: 1) time-domain loss, denoted as the waveform loss, that measures the distortion between the natural and generated waveform; and 2) frequency-domain loss, that measures the Mel-scale acoustic feature loss between the natural and generated acoustic features. WaveTTS ensures both the quality of the acoustic features and the resulting speech waveform. To our best knowledge, this is the first implementation of Tacotron with joint time-frequency domain loss. Experimental results show that the proposed framework outperforms the baselines and achieves high-quality synthesized speech. 

\end{abstract}

\section{Introduction}


In recent years, Text-to-Speech (TTS) technology has advanced by leaps and bounds, from concatenative approach \cite{hunt1996unit,ze2013statistical,merritt2016deep} to statistical parametric approach \cite{zen2009statistical,tokuda2013speech,zen2015unidirectional,wu2016investigating}, and to deep learning \cite{wang2017tacotron}.  They produce high quality and natural speech that rival human vocal production \cite{taylor2009text,zen2009statistical}. TTS is also widely used in human-machine communications, such as robotics, call centres, games, entertainments, and healthcare applications.

With the advent of deep learning, neural approaches to TTS become mainstream, such as Tacotron \cite{wang2017tacotron}, Tacotron2 \cite{shen2018natural} and its varieties \cite{skerry2018towards,hsu2018hierarchical,habib2019semi,liu2017mongolian,liu2019teacher,li2018end}.
The end-to-end TTS model is based on encoder-decoder framework, which has been widely adopted for sequence generation tasks, such as speech recognition \cite{graves2014towards,bahdanau2016end,amodei2016deep,chen2019end}, image translation \cite{ak2019attribute} and neural machine translation \cite{bahdanau2014neural,johnson2017google}. Tacotron-based TTS typically consists of two modules: 1) feature prediction, and 2) waveform generation. The main task of feature prediction network is to obtain frequency-domain acoustic features, while the waveform generation module is to convert frequency-domain acoustic features into time-domain waveform. 

A typical Tacotron implementation adopts Griffin-Lim algorithm \cite{griffin1984signal,masuyama2019deep} for phase reconstruction, that only uses a loss function derived from amplitude spectrogram in frequency domain. Such a loss function doesn't take the resulting waveform into consideration in the optimization process. As a result, there exists a mismatch between the Tacotron optimization and the expected waveform.
We note that such mismatch also exists in many other speech processing tasks, such as speech separation \cite{wang2018supervised}, 
where we observe that, by incorporating time-domain loss function \cite{wang2015deep}, one can improve the output speech quality.
More recently, deep learning approach to speech enhancement methods with time-domain raw waveform outputs \cite{fu2017raw,liu2019multichannel} have also been investigated. However, we note that time-domain loss function has not been well explored in speech synthesis, which will be the focus of this paper. 

Tacotron2 \cite{shen2018natural} has been proposed to achieve high quality synthesized voice. It addresses the waveform optimization problem by using WaveNet-based neural vocoder \cite{oord2016wavenet,berrak-journal,sisman2018adaptive,berrak_is18,hayashi2017investigation}. We note that WaveNet avoids the artifacts and deterioration caused by deterministic vocoders. It generates time-domain waveform samples conditioned on the predicted mel-spectrum features. Although Tacotron2 allows end-to-end learning of TTS directly from character sequences and speech waveforms, its feature prediction network is trained independently of the WaveNet vocoder.  At run-time, the feature prediction network and WaveNet vocoder are artificially joined together. As a result, the framework suffers from the mismatch between frequency-domain acoustic features and time-domain waveform. It is reported that the samples generated from WaveNet occasionally become unstable, especially when less accurately predicted acoustic features are used as the local condition parameters \cite{sisman2019machine, tobing2019voice}. To overcome such mismatch, we propose to use joint time-frequency domain loss for TTS that effectively improves the synthesized voice quality.

In this paper, we propose to add a time-domain loss function to the Griffin-Lim/ISTFT output of Tacotron-based TTS model at the training time. 
In other words, we use both frequency-domain loss and time-domain loss for the training of feature prediction model. We hypothesize that the feature prediction network will compensate the possible artifacts that Griffin-Lim process may introduce under the supervision of the time-domain loss. We use Griffin-Lim iteration followed by ISTFT to transform frequency-domain feature to time-domain waveform and use scale-invariant signal-to-distortion (SI-SDR) \cite{le2019sdr,kolbaek2019loss} to measure the quality of the time-domain waveform. Our proposed idea shares a similar motivation with \cite{zhao2018wasserstein} in terms of the use of waveform loss. However, it differs from \cite{zhao2018wasserstein} in many ways, for example, we study Tacotron-based TTS, while \cite{zhao2018wasserstein} mostly deals with Wasserstein GAN-based TTS. 

The main contributions of this paper include: 1) we study the use of time-domain loss for speech synthesis; 2) we improve  Tacotron-based TTS framework by proposing a new training scheme based on joint time-frequency domain loss; and
3) we propose to use SI-SDR metric to measure the distortion of time-domain waveform. The novel training scheme optimizes the frequency-domain acoustic features in a way that it leads to better time-domain waveform. To our best knowledge, this is the first implementation of joint training scheme on frequency and time domain for Tacotron-based TTS framework.

This paper is organized as follows:  In Section 2, we present the Tacotron-based baseline TTS system. In Section 3, we present the novel idea of joint time-frequency domain loss, and formulate the training and run-time processes. We report the experimental results in Section 4. Section 5 concludes the study.

\section{Baseline: Tacotron-based TTS}
\label{sec:baseline}

\begin{figure}[t]
  \centering
  \centerline{\includegraphics[width=9cm]{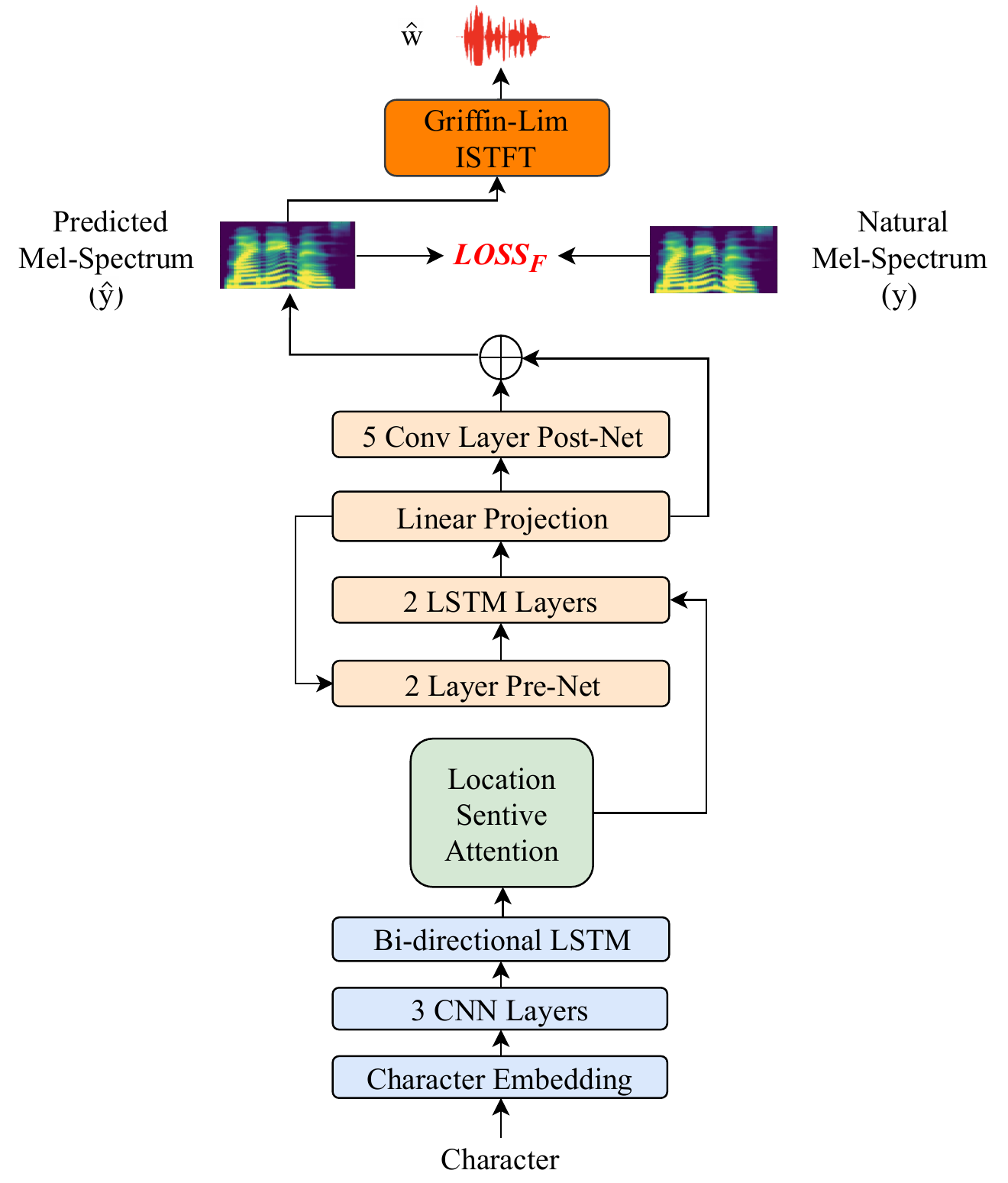}}
\vspace{-5mm}
\caption{Block diagram of Tacotron-based reference baseline that has three modules, encoder, attention-based decoder, and  Griffin-Lim reconstruction algorithm.}
\label{fig:baseline}
\vspace{-5mm}
\end{figure}

In this paper, we use a Tacotron-based framework \cite{shen2018natural} as a reference baseline. 
We illustrate the overall architecture of the \textit{reference baseline} in Figure \ref{fig:baseline}, that includes feature prediction model which contains encoder, attention-based decoder and Griffin-Lim algorithm for waveform reconstruction. The encoder (blue box in Figure \ref{fig:baseline}) consists of two components, a CNN-based module that has 3 convolution layers, and a LSTM-based module that has a bidirectional LSTM layer \cite{liu2019imu,liu2018improving,liu2018lstm,liu2018phonologically,liu2019building}. The decoder (pale yellow box in Figure \ref{fig:baseline}) consists of four components: a 2-layer pre-net, 2 LSTM layers, a linear projection layer and a 5-convolution-layer post-net. The decoder is a standard autoregressive recurrent neural network that generates the mel-spectrum features and stop tokens frame by frame. 

During training, we optimize the feature prediction model to minimized the frequency-domain loss between the generated mel-spectrum features ($\hat y$) and the target mel-spectrum features ($y$). 







\vspace{-4mm}
\section{WaveTTS}

In this section, we study the use of a newly proposed time-domain loss function for Tacotron-based TTS. By applying a new training strategy that takes into account both time and frequency domain loss functions, we effectively reduce the mismatch between the frequency-domain features and the time-domain waveform, and improve the output speech quality. In addition, Griffin-Lim algorithm and SI-SDR metric are utilized to realize the calculation process of proposed loss term. The proposed framework is called as \textit{WaveTTS} hereafter. 

\subsection{Time-domain and Frequency-domain Loss Functions}
\begin{figure}[t]
  \centering
  \centerline{\includegraphics[width=9cm]{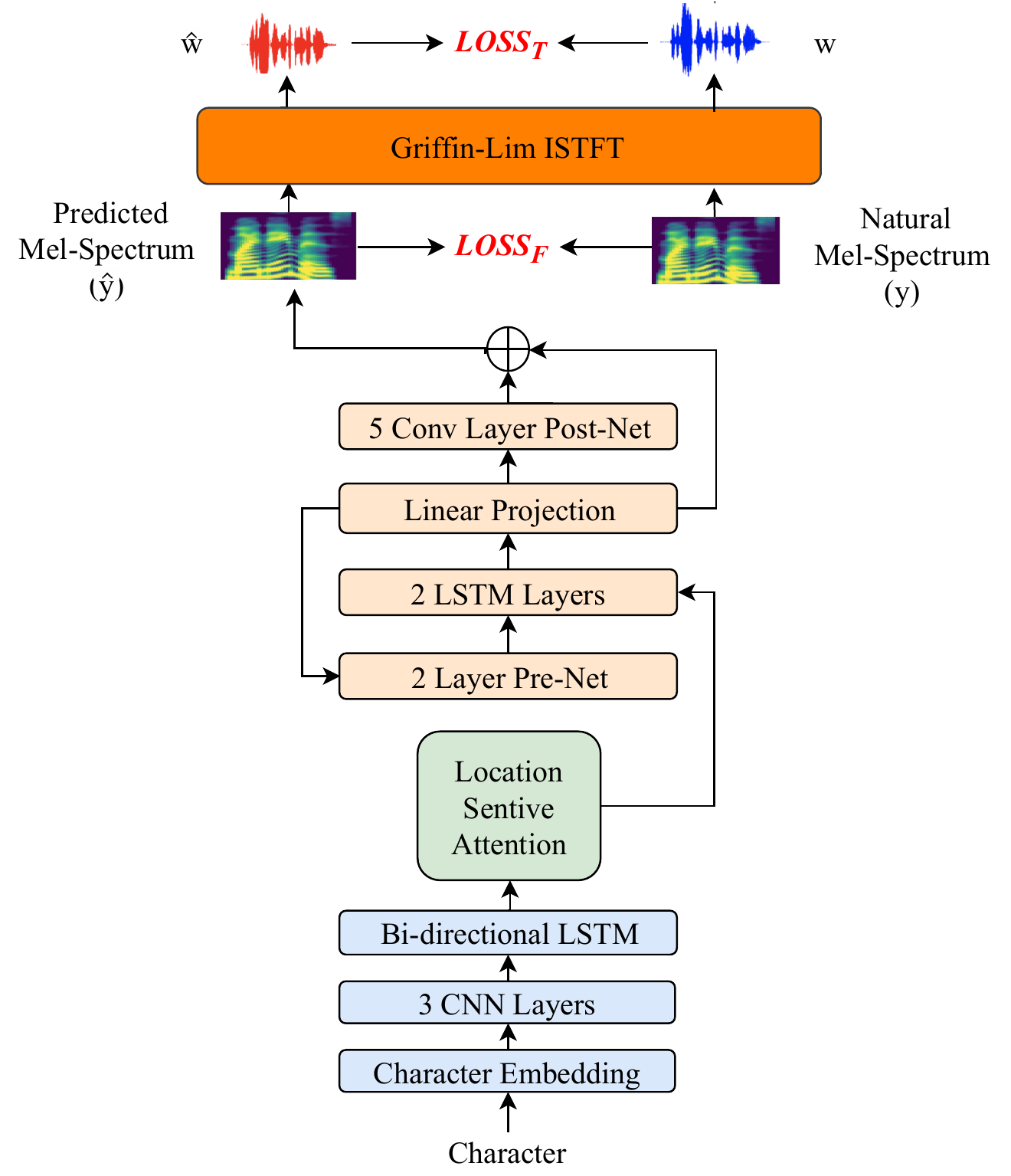}}
  \vspace{-4mm}
\caption{Block diagram of proposed WaveTTS model using both frequency-domain loss ($Loss_{F}$) and time-domain loss ($Loss_{T}$).}
\vspace{-6mm}
\label{fig:proposed}

\end{figure}

In WaveTTS, we define two objective functions during training: 1) frequency-domain loss, denoted as $Loss_{F}$, that is calculated with the mel-spectrum features in a similar way described in \cite{shen2018natural}; and 2) the proposed time-domain loss, denoted as $Loss_{T}$, that is obtained at waveform level at the output of Griffin-Lim iteration, that estimates time-domain signal from the mel-spectrum features. The two objective functions are illustrated in Figure \ref{fig:proposed}.



The entire process can be formulated as follows. The encoder takes the character sequence $x = (x_{1},x_{2},...,x_{T})$ as input and converts the one-hot vector to continuous features representation $h$:
\begin{equation}
h_{t} = {\rm Encoder}(h_{t-1},x_{t})
\end{equation}

The decoder outputs a mel-spectrum feature $\hat y_t$ at each step $t$:
\begin{equation}
\hat y_{t} = {\rm Decoder}(y_{t-1}, \sigma(h_{t}))
\end{equation}
where $\sigma()$ represents a function  to calculate the context vector by using location-sensitive attention mechanism \cite{vaswani2017attention}.

We first calculate $Loss_{F}$ in a similar way as that in \cite{shen2018natural}. $Loss_{F}$ ensures that the generated mel-spectrum is close the natural mel-spectrum. $Loss_{F}$ is given as follows,

\vspace{-1mm}
\begin{equation}
Loss_{F} (\hat{y},y) = \sum_{t=1}^{T'} L_{2} (\hat{y}_{t},y_{t})
\end{equation}

\noindent{where} $T'$ is the total number of the sequences in training data. $L_{2}$ loss function is used to minimize the error which is the sum of the all the squared differences between the true value and the predicted value.

We then propose the use of a time-domain loss $Loss_{T}$, that is applied to the Griffin-Lim output. This is to reduce the mismatch between the optimized frequency-domain output and the actual time-domain waveform \cite{zhao2018wasserstein}. The implementation details of $Loss_{T}$ will be explained in Section 3.2. 

Overall, the proposed WaveTTS framework has two loss functions. $Loss_{F}$ minimizes the loss between converted and original mel-spectrum, while $Loss_{T}$ minimizes this loss at waveform level. We add a weighting coefficient $\lambda$ to balance the two losses. The training criterion of the whole model is defined as:
 \vspace{-1mm}
\begin{equation}
{Loss= Loss_{F} + \lambda \cdot Loss_{T}}
\label{eqt:loss}
\end{equation}


Algorithm 1 also shows the complete training process of our proposed WaveTTS.
WaveTTS model predicts the mel-spectrum features $\hat y$ from the given input character sequence $x = (x_{1}, x_{2}, ..., x_{T})$, and then converts the estimated and target mel-spectrum to the time-domain signal $\hat w$, $w$ using Griffin-Lim based ISTFT algorithm (blue content in Algorithm 1). Finally, the joint loss function given in Equation \ref{eqt:loss} is used to optimize the WaveTTS model.

\subsection{Implementation of time-domain Loss}

\subsubsection{Time-domain loss}
We adopt Griffin-Lim algorithm \cite{griffin1984signal}, followed by ISTFT to generate the time-domain waveform. Griffin-Lim algorithm has been widely used in speech synthesis \cite{shen2018natural, tjandra2019vqvae} for its simplicity, that can be formulated as follows: 


\vspace{-1mm}
\begin{equation}
\hat A = {\varepsilon  }(\hat y)
\end{equation}

\vspace{-3mm}
\begin{equation}
 \mathbf{ \hat X} = {\mathrm{Griffin\!\!-\!\!Lim} }(\hat A)
\end{equation}


\noindent{where} $\hat y$
represents the predicted mel-spectrum sequences, and ${\hat A}$ represents their amplitude. $\varepsilon $ is the function that calculates the amplitude of the given input mel-spectrum sequences, which is followed by the Griffin-Lim algorithm, that estimates a complex valued spectrum, while minimizing the change to the input amplitude ${\hat A}$. ISTFT transforms the estimated complex valued spectrum to time-domain signals. The details of the Griffin-Lim algorithm are given in Algorithm 2. $P_{\mathcal{S}}$ is the metric projection onto a set $\mathcal{S}$. Here, $\mathcal{C}$ is the set of consistent spectrums, and $\mathcal{A}$ is the set of spectrums whose amplitude is the same as the given one.

It's worth mentioning that the Griffin-Lim algorithm usually requires many iterations, as shown in Algorithm 2, at run-time to obtain a high-quality audio signal. It is an optimization process independent of Tacotron training. We would like the Tacotron feature prediction network to generate acoustic features, that not only are close to those of natural speech in frequency-domain, but also allow Griffin-Lim to produce speech that is close to natural speech in time-domain.

Let's denote the predicted and the original mel-spectrum as ($\hat y, y$). We apply Griffin-Lim and ISTFT to generate ($\hat w, w$), with $\hat y \rightarrow \hat w$ and $y \rightarrow w$. We keep the same number of Griffin-Lim iterations to ensure that Griffin-Lim behaves the same between the two transformation pairs, $\hat y \rightarrow \hat w$ and $y \rightarrow w$. We measure the distortion between $\hat w$ and $w$ with a time-domain loss, that forces the speech waveform generated from the predicted network to be as close as possible to that generated from the mel-spectrum of natural speech.




 \begin{algorithm}[t]
\label{algo:main}
\setstretch{1.1}
\SetAlgoLined
\textbf{Input}:\\
\quad \quad Training set: $D = \{x ,y ,w \} $\\
\quad \quad $x $: character sequence\\
\quad \quad $y $: frequency-domain acoustic feature sequence\\

\textbf{Output}:\\
\quad \quad $\varTheta$: TTS model

\textbf{Hyper-parameter}:\\
\quad \quad N: epoch number \\
\quad \quad n: batch size  \\
\quad \quad $\eta$: learning rate \\
\quad \quad $\lambda$: weight for time-domain loss \\

\textbf{Begin}
\\1: Initialize TTS model $\varTheta$
\\2: \textbf{for} epoch = 1,2,...N \textbf{do}
\\3: \quad \textbf{for} batch = 1,2,...n \textbf{do}
\\4: \qquad output $\hat{y}$:
\\ \qquad \qquad $\hat{y} = \varTheta(x)$
\\5: \qquad generate $\hat{w}$ and $w$ using the Griffin-Lim \\ \qquad \quad algorithm:
\\ \qquad \qquad $\hat{A} = \varepsilon (\hat{y})$
\\ \qquad \qquad ${A} = \varepsilon ({y})$
\\ \qquad \qquad $\textcolor{blue}{\mathbf{\hat{X}} = \mathrm{Griffin\!\!-\!\!Lim}(\hat{A})}$
\\ \qquad \qquad $\textcolor{blue}{\mathbf{{X}} = \mathrm{Griffin\!\!-\!\!Lim}({A})}$
\\ \qquad \qquad $\textcolor{blue}{\hat w = {\rm ISTFT}( \mathbf{\hat{X}}  )}$
\\ \qquad \qquad $ \textcolor{blue}{w  = { \rm ISTFT}({  \mathbf{X}  })}$
\\6: \qquad update $\varTheta$ with $Loss_{F}$ and $Loss_{T}$ :
\\ \qquad  \quad  $\varTheta \leftarrow \bigtriangledown_{\varTheta}(\textcolor{blue}{Loss_{F}(y,\hat{y})+\lambda \cdot Loss_{T}(w,\hat{w})})$
\\7: \quad \textbf{end for}
\\8: \textbf{end for}
\\9: return $\varTheta$
\\\textbf{End}

 \caption{Training WaveTTS model using time-frequency domain loss.}
\end{algorithm}

\subsubsection{Scale-Invariant Signal-to-Distortion (SI-SDR)}

In speech synthesis, we optimize the feature prediction network to minimize the discrepancy between the synthesized waveform and the target natural speech that is supervised by a loss function. We propose a time-domain loss function, $Loss_{T}$, that is based on scale-invariant signal-to-distortion (SI-SDR). 
SI-SDR has been introduced as a time-domain objective measure in source separation \cite{luo2018tasnet,bahmaninezhad2019comprehensive,venkataramani2018performance} to compare two time-domain speech signals. We adopt SI-SDR to measure the discrepancy between the generated waveform and the target natural speech. To our best knowledge, this is the first implementation of SI-SDR for time-domain loss calculation to improve TTS quality.

We note that SI-SDR is evaluated only during training, and not required at run-time inference. During training, the predicted time-domain waveform $\hat w$ and the target speech $w$ have identical duration. Similarly, the predicted mel-spectrum $\hat y$ and target mel-spectrum $y$ also share the same frame length, that facilitates the SI-SDR calculation. As a greater SI-SDR indicates better quality, to turn it into a loss function, we take the negative value of SI-SDR as the loss function,


\begin{equation}
\begin{aligned} 
Loss_{T}(\hat w, w)= -10 \log _{10} \frac{\|\alpha w\|^{2}}{\|\alpha w-\hat{w}\|^{2}} \end{aligned}
\end{equation}
where
\begin{equation}
\alpha=\frac{\hat{w}^{T} w}{\|w\|^{2}}=\underset{\alpha}{\operatorname{argmin}}\|\alpha w -\hat{ w }\|^{2}
\end{equation}
SI-SDR is expressed in decibel (dB)
and defined in the range of $[-\infty,\infty]$, so is $Loss_{T}$.




\begin{table*}[]
\centering
\label{table:system}

\begin{tabular}{|c|c|c|c|c|}
\hline
\multicolumn{2}{|c|}{\multirow{2}{*}{}} & \multicolumn{2}{c|}{\textbf{Training Phase}}                          & \textbf{Run-time Inference }           \\ \cline{3-5} 
\multicolumn{2}{|c|}{}                  & \textbf{Loss Function}                    & \textbf{Waveform Generation} & \textbf{Waveform Generation} \\ \hline
\multirow{2}{*}{\textit{Baseline}} & \textbf{Tacotron-GL} & $ Loss_{F}$            & NA                  & Griffin-Lim  \cite{griffin1984signal}       \\ \cline{2-5} 
                          & \textbf{Tacotron-WN} & $ Loss_{F}$          & NA                  & WaveNet vocoder \cite{hayashi2017investigation}     \\ \hline
\multirow{2}{*}{\textit{Proposed}} & \textbf{WaveTTS-GL}  & $ Loss_{F} + \lambda \cdot Loss_{T} $ & Griffin-Lim  \cite{griffin1984signal}       & Griffin-Lim \cite{griffin1984signal}        \\ \cline{2-5} 
                          & \textbf{WaveTTS-WN}  & $ Loss_{F} + \lambda \cdot Loss_{T} $  & Griffin-Lim   \cite{griffin1984signal}      & WaveNet vocoder \cite{hayashi2017investigation}    \\ \hline
\end{tabular}
\caption{Comparison of the frameworks in terms of loss function and waveform generation during training and run-time.}
\end{table*}

\begin{algorithm}[t]

\label{algo:main}
\setstretch{1.2}
\SetAlgoLined
\textbf{Input}: \\
\quad \quad $\angle c_{0}$: the initial phase \\
\quad \quad $ A$: amplitude

\textbf{Output}:\\
\quad \quad $\mathbf{X}$: complex valued spectrum
\\
\textbf{Begin}
\\1: Initialize $\mathbf{X}_{0}=A \cdot e^{\cdot i \angle c_{0}}$
\\2: \textbf{for} iteration = 1,2,...n \textbf{do}
\\ \quad \quad$\mathbf{X}_{n}=P_{\mathcal{C}}(P_{\mathcal{A}}(\mathbf{X}_{n-1}))$
\\3: \textbf{end for}
\\4: return  $\mathbf{X}_{n}$
\\\textbf{End}

\caption{Griffin-Lim algorithm.}
 
\end{algorithm}

\vspace{-2mm}
\section{Experiments}
We report the TTS experiments on LJSpeech database \footnote{https://keithito.com/LJ-Speech-Dataset/}, which consists of 13,100 short clips with a total of nearly 24 hours of speech from one single speaker reading about 7 non-fiction books. We develop four systems for a comparative study: 
\begin{itemize}
    \item \textit{Tacotron-GL}: Tacotron-based baseline model \cite{shen2018natural}, that has only frequency-domain loss function. Griffin-Lim algorithm is used to generate the waveform at run-time.
    \item \textit{Tacotron-WN}: Tacotron-based baseline model \cite{shen2018natural}, that has only frequency-domain loss function.  Pre-trained WaveNet vocoder is used to generate the waveform at run-time. 
    \item \textit{WaveTTS-GL}: proposed WaveTTS model is trained with joint time-frequency domain loss. Griffin-Lim algorithm is used during training and run-time phases. 
    \item \textit{WaveTTS-WN}: proposed WaveTTS model is trained with joint time-frequency domain loss. Griffin-Lim algorithm is used during training and the pre-trained WaveNet vocoder is used to synthesize speech at run-time. 
\end{itemize}


\vspace{-2mm}
We also compare these systems with the ground truth speech, denoted as \textit{GT}. The comparison of the systems is also summarized in Table 1.
\vspace{-1mm}

\subsection{Experimental Setup}
The 80-channel mel-spectrum is extracted with 12.5ms frame shift and 50ms frame length. It is normalized to zero-mean and unit-variance as the reference target. The decoder predicts only one non-overlapping output frame at each decoding step. We use the Adam optimizer with $\beta_1$ = 0.9, $\beta_2$ = 0.999 and a learning rate of $10^{-3}$ exponentially decaying to $10^{-5}$ starting with 50k iterations. We also apply $L_{2}$ regularization with weight $10^{-6}$. Hyper-parameter $\lambda$ in Equation \ref{eqt:loss} is empirically set as $10^{-3}$. All models are trained with a batch size of 32. The final models are trained with 100k steps for all systems. 
At run-time, Tacotron-GL and WaveTTS-GL use Griffin-Lim algorithm with 64 iterations, while Tacotron-WN and WaveTTS-WN use pretrained WaveNet vocoder. 



\subsection{Subjective Evaluation}

\begin{figure}[t]

  \centering
  \centerline{\includegraphics[width=65mm]{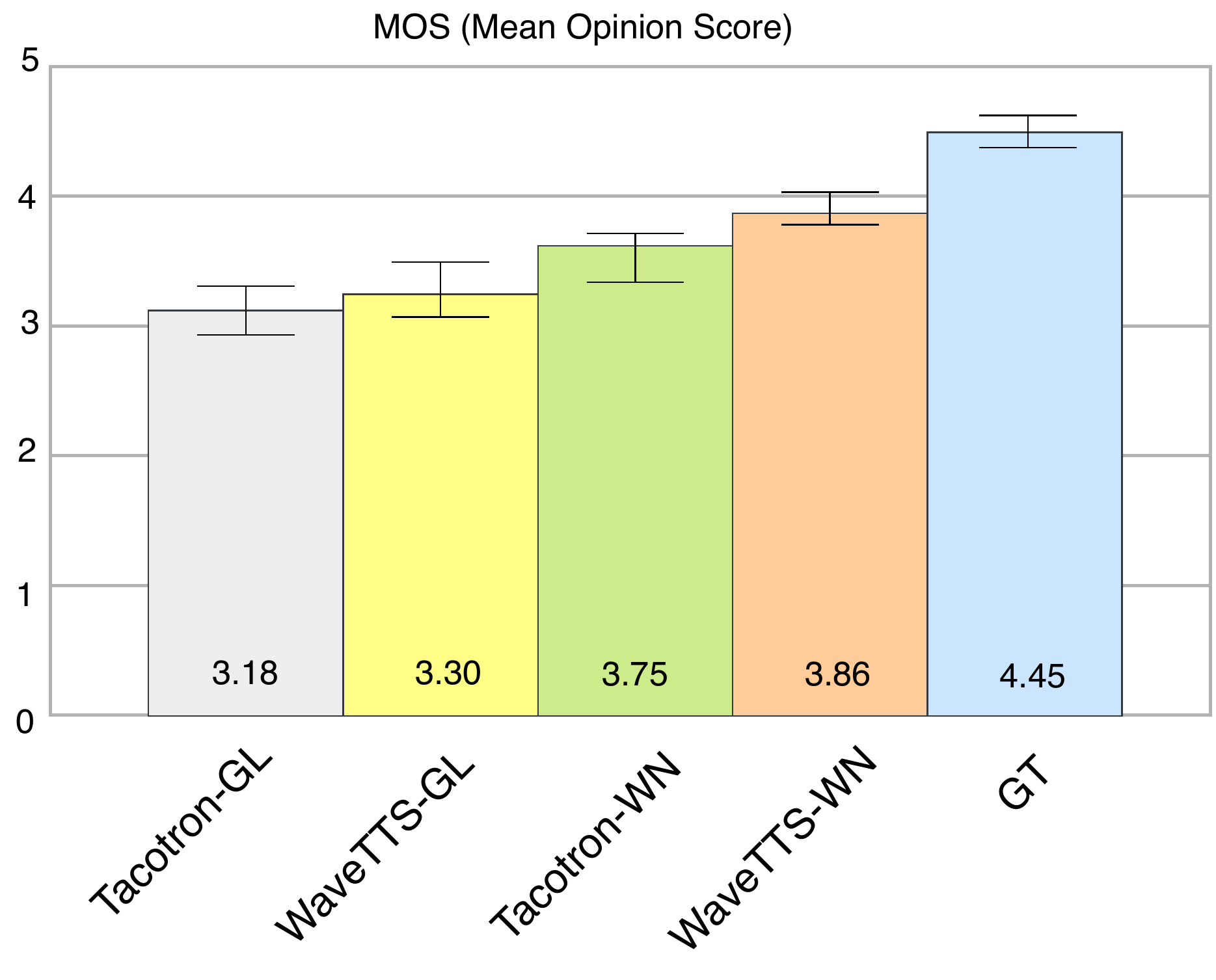}}
   \vspace{-5mm}
\caption{Comparison of mean opinion scores (MOS) between Tacotron-GL, WaveTTS-GL, Tacotron-WN, WaveTTS and ground truth (GT), with 95\% confidence interval. }
\vspace{-3mm}
\label{fig:mos}
\end{figure}

We conduct listening experiments to evaluate the quality of the synthesized speech. We first evaluate the sound quality of the synthesized speech in terms of mean opinion score (MOS) among GT, Tacotron-GL, Tacotron-WN and the proposed WaveTTS-GL and WaveTTS-WN, that is reported in Figure \ref{fig:mos}. 

The listeners rate of the quality is on a 5-point scale: “5” for excellent, “4” for good, “3” for fair, “2” for poor, and “1” for bad. The MOS values reported in Figure \ref{fig:mos} are calculated by taking the arithmetic average of all scores assigned the subjects who have passed the validation question test. We keep the linguistic content the same among different models so as to exclude other interference factors. 15 subjects  participate in these experiments, and each one of them listens to 120 synthesized speech samples. We have three observations through the experiments: 
\begin{enumerate}
    \item The importance of joint time-frequency domain loss: We compare Tacotron-GL and WaveTTS-GL to observe the effect of joint time-frequency domain loss. We believe that this is a fair comparison as both frameworks use Griffin-Lim algorithm for waveform generation during training and/or run-time. As can be seen in Figure \ref{fig:mos}, WaveTTS-GL outperforms Tacotron-GL by achieving 3.30 MOS value, while Tacotron-GL achieves only 3.18.  
    \item The performance of WaveTTS with a neural vocoder at run-time: We compare Tacotron-WN and WaveTTS-WN to investigate how well the predicted mel-spectrum features perform with WaveNet vocoder. We observe that even though WaveTTS is trained with Griffin-Lim algorithm, it performs better than Tacotron when WaveNet vocoder is available at run-time. This shows how well our proposed WaveTTS performs with other neural vocoders.
    \item Griffin-Lim vs WaveNet vocoder at run-time: We compare WaveTTS-GL and WaveTTS-WN in terms of voice quality. We note that both frameworks are trained under the same conditions. However, WaveTTS-WN uses WaveNet vocoder for waveform generation at run-time. As expected, WaveTTS-WN outperforms WaveTTS-GL.  
\end{enumerate}

\begin{figure}[t]

  \centering
  \centerline{\includegraphics[width=70mm]{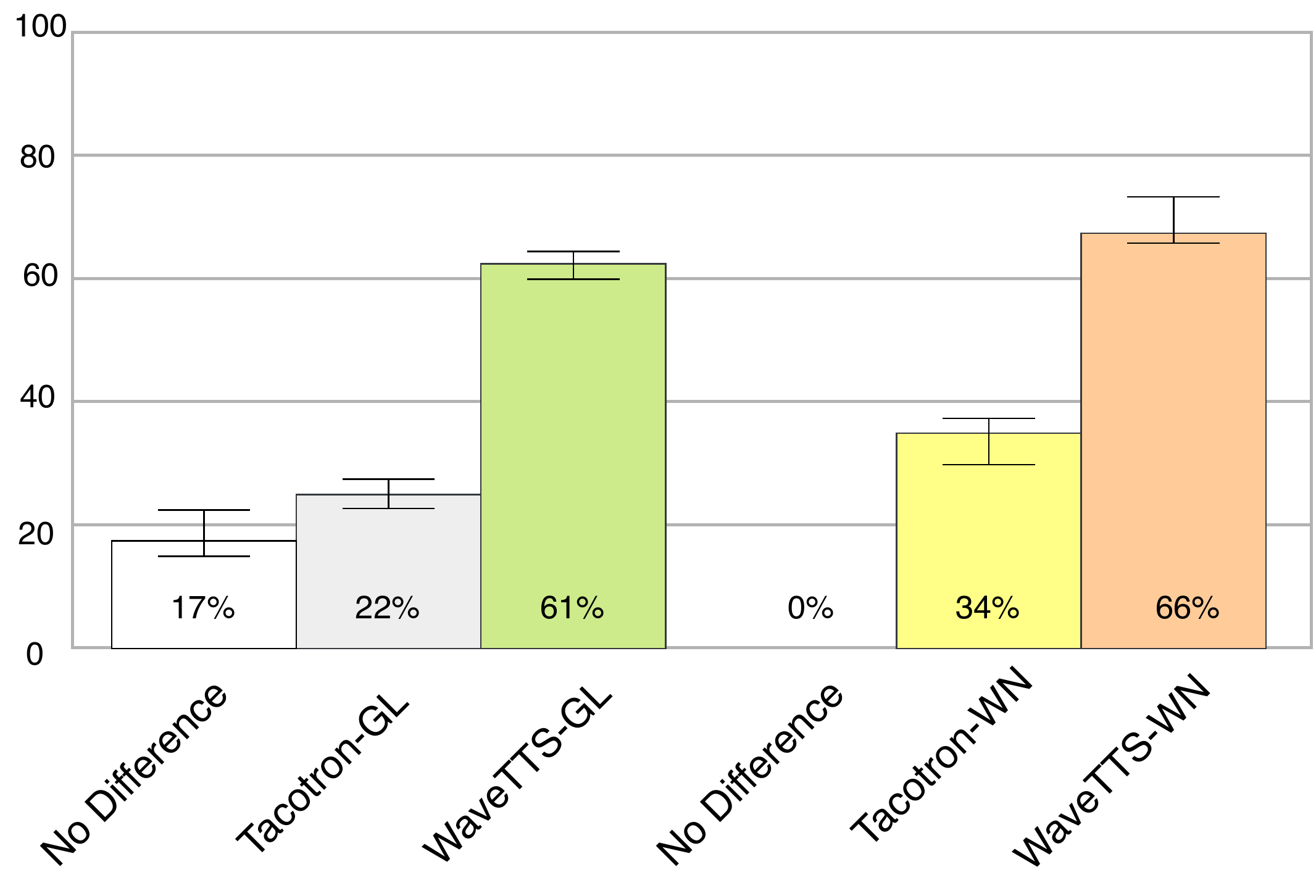}}
  \vspace{-5mm}
\caption{The preference test between (1) Tacotron-GL and  WaveTTS-GL, and (2) Tacotron-WN and WaveTTS-WN, with 95\% confidence interval.}
\label{fig:ab}
\vspace{-4mm}
\end{figure}
\begin{figure}

  \centering
  \centerline{\includegraphics[width=55mm]{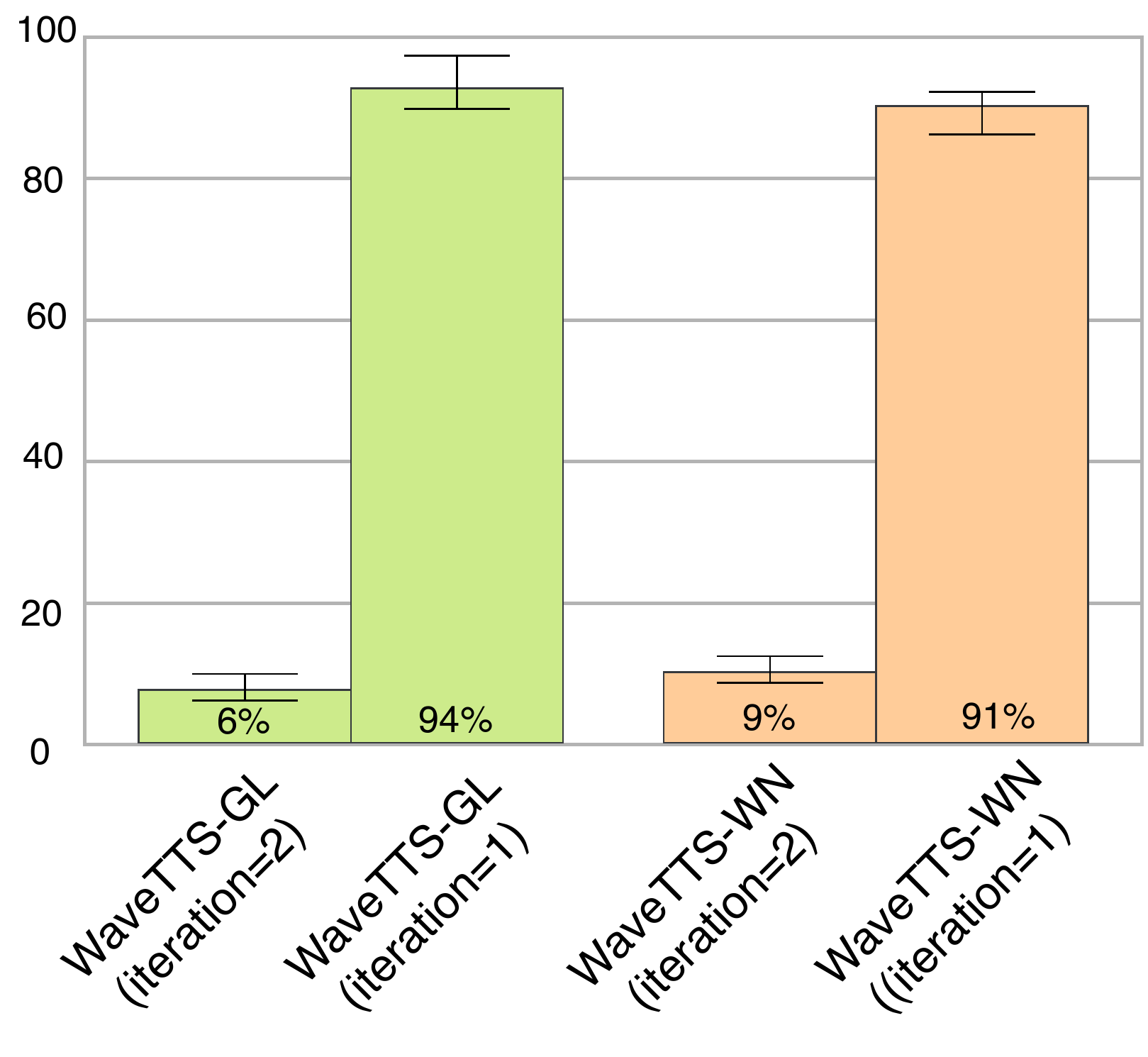}}
  \vspace{-5mm}
\caption{The preference test between (1) WaveTTS-GL (2 iterations) and  WaveTTS-GL (1 iteration), and (2) WaveTTS-WN (2 iterations) and  WaveTTS-WN (1 iteration), with 95\% confidence interval.}
\label{fig:gl-ab}
 
\end{figure}

We also conduct A/B preference tests to assess speech quality of proposed frameworks. In A/B preference tests, the listeners are asked to compare the quality and naturalness of the synthesized speech samples from different systems, and select the better one. 15 listeners were invited to participate in all the tests. 80 samples were randomly selected from 200 converted samples from each system.  Figure \ref{fig:ab} shows the speech quality test results, which suggests that our proposed WaveTTS framework outperforms the baseline system for both Griffin-Lim and WaveNet vocoder settings at run-time. 
 
We further conduct another A/B preference test to examine the effect of the number of Griffin-Lim iterations on the WaveTTS performance. To calculate the time-domain loss, WaveTTS needs to generate the synthesized waveform during training. For rapid turn-around, we only apply 1 and 2 Griffin-Lim iterations for phase reconstruction, and investigate the effect in terms of voice quality. Figure \ref{fig:gl-ab} shows A/B preference test results on both WaveTTS-GL and WaveTTS-WN. We observe that the single iteration of Griffin-Lim algorithm presents a better performance than 2 iterations.

\section{Conclusion}
In this paper, we propose a new Tacotron implementation, called WaveTTS. The traditional framework calculates only frequency-domain loss to update the network parameters, that doesn't directly control the quality of the generated time-domain waveform. The proposed WaveTTS is unique in a sense that it calculates both time-domain and frequency-domain loss, and optimizes the model for generating high-quality synthesized voice. We propose to use scale-invariant signal-to-distortion (SI-SDR) as the loss function. Even though the proposed model is trained with Griffin-Lim algorithm for time-domain loss calculation, it performs remarkable well with both Griffin-Lim and WaveNet vocoder at run-time. Experimental results show that the proposed framework outperforms the baselines and achieves high-quality synthesized speech. To our best knowledge, this is the first implementation of Tacotron-based TTS model with joint time-frequency domain loss.

As a future work, we will investigate the training phase of joint time-frequency domain loss with a neural vocoder for high-quality TTS. 

\section{Acknowledgements}
This research was supports by the National Natural Science Foundation of China (No.61563040, No.61773224), Natural Science Foundation of Inner Mongolian (No.2018MS06006, No.2016ZD06). This work is also supported by Human-Robot Interaction Phase 1 (Grant No. 192 25 00054), National Research Foundation Singapore under the National Robotics Programme. It is also supported by National Research Foundation Singapore under the AI Singapore Programme (Award Number: AISG-100E-2018-006), and Programmatic Grant No. A18A2b0046 (Human Robot Collaborative AI for AME) and A1687b0033 (Neuromorphic Computing) from the Singapore Government’s Research, Innovation and Enterprise 2020 plan in the Advanced Manufacturing and Engineering domain.


\bibliographystyle{IEEEbib}
\bibliography{Odyssey2020_BibEntries}

%

\end{document}